# AU-SN FLIP-CHIP SOLDER BUMP FOR MICROELECTRONIC AND OPTOELECTRONIC APPLICATIONS

*Jeong-Won Yoon, Hyun-Suk Chun, Ja-Myeong Koo and Seung-Boo Jung*

School of Advanced Materials Science and Engineering, Sungkyunkwan University
300 Cheoncheon-dong, Jangan-gu, Suwon, Gyeonggi-do 440-746, Korea

## ABSTRACT

As an alternative to the time-consuming solder pre-forms and pastes currently used, a co-electroplating method of eutectic Au-Sn alloy was used in this study. Using a co-electroplating process, it was possible to plate the Au-Sn solder directly onto a wafer at or near the eutectic composition from a single solution. Two distinct phases, $Au_5Sn$ and $AuSn$, were deposited at a composition of 30at.%Sn. The Au-Sn flip-chip joints were formed at 300 and 400℃ without using any flux. In the case where the samples were reflowed at 300℃, only an $(Au,Ni)_3Sn_2$ IMC layer formed at the interface between the Au-Sn solder and Ni UBM. On the other hand, two IMC layers, $(Au,Ni)_3Sn_2$ and $(Au,Ni)_3Sn$, were found at the interfaces of the samples reflowed at 400℃. As the reflow time increased, the thickness of the $(Au,Ni)_3Sn_2$ and $(Au,Ni)_3Sn$ IMC layers formed at the interface increased and the eutectic lamellae in the bulk solder coarsened.

## 1. INTRODUCTION

With the rapid advancement of information technologies which has been seen in recent years, the use of optoelectronic packages, such as planar light-wave circuit modules, is increasing rapidly. In these packages, solder alloys are commonly employed for the purpose of mounting active devices, such as laser diodes, on the substrate of the package [1]. The solder interconnection of the module not only provides conventional functions, such as heat dissipation, electrical connection and self-aligning effects, but also helps to maintain the precise alignment between the laser diode and the waveguide during the use of the module. The solders used for the application of these modules, therefore, must be resistant to the creep deformation induced by stresses, such as thermal stress, in the modules.

Solders for bonding applications in microelectronic/optoelectronic packages are classified as either soft solder or hard solder depending on their melting temperature [2]. Soft solders, such as Sn and In alloys, have low melting temperatures, but exhibit lower yield strengths, which lead to lower creep resistance. Solder creep reduces the reliability of optoelectronic packages, because the alignment of the devices cannot be maintained over time. On the other hand, hard solders, including Au-rich Au-Sn, Au-Si, and Au-Ge alloys, have higher melting temperatures and higher yield strengths. The advantages of solders with a higher melting temperature include superior thermal stability and long term reliability. Among these hard solders, eutectic Au-30at.% Sn is the preferred alloy, because of its relatively low melting point, low elastic modulus, high thermal conductivity and high strength, as compared with those of the other solders [3-5]. Au-Sn eutectic solder has a lower melting temperature (278°C) compared to other hard solders, such as Au-Si (363°C) and Au-Ge (356°C). This property makes it useful for bonding devices that are sensitive to high processing temperatures, but need good creep resistance, such as GaAs or large Si dies on alumina. In addition, the high thermal conductivity of Au-Sn (57 W/m℃) makes it particularly useful for bonding higher power devices that demand good heat dissipation. Au-Sn solder also offers many other advantages when making solder joints, such as the ability to solder without using flux, the formation of a hermetic seal, good mechanical and electrical properties and low intermetallic growth rates when used over Ni, Pd or Pt. The drawback of this solder is that maintaining the desired eutectic composition requires extreme accuracy and precise control.

Au-30at.% Sn solder has traditionally been applied using solder pre-forms, paste, or electron-beam evaporation [3]. Solder pre-forms are problematic due to the alignment difficulties that are encountered and the oxidation of the solder prior to bonding. Solder paste also suffers from oxidation prior to bonding, in addition to the possibility of solder contamination during bonding originating from the organic binder in the paste. Electron-beam evaporation is advantageous for Au-Sn solder deposition in that the amount of oxide formed prior to bonding can be reduced and the thickness and position of the solder can be precisely controlled. The sequential evaporation of Au and Sn layers to produce a deposit of the desired composition has been successfully employed, along with co-evaporation techniques.

The electroplating of eutectic Au-Sn solder is also an attractive alternative, because it offers the advantages of low cost and high speed as compared to evaporation techniques, while providing a similar level of control to that of pre-forms and paste. Au–Sn solder has been deposited by electroplating Au and Sn layers sequentially from separate Au and Sn solutions. Recently, however, Au–Sn solder has been co-deposited by electroplating [5]. The alloy electroplating of Au-Sn promises better composition control, lower mechanical stress, and finer dimensional capability, along with lower processing complexity, higher throughput, and a lower capital cost. Co-deposition is also advantageous in that the level of Sn oxidation is kept to a minimum during the electroplating process, since the wafer does not need to be removed from the solution until plating is complete.

Amongst the major trends observed in the development of semiconductor devices is the small volume of the products that was achieved using integrated circuits (ICs) and the larger size and functionality per unit area of the modules used in the products. To facilitate these trends, flip chip (FC) interconnection technology was developed. An FC interconnection is the connection of an IC chip to a carrier or substrate with the active face of the chip facing toward the substrate. FC technology is generally considered as the ultimate first level connection, because it allows for the highest density and shortest path length to be achieved, so that the optimal electrical characteristics can be obtained.

Flip-chip solder connections have to be fluxless, because flux residue seriously affects both the performance and reliability of flip-chip assemblies [6]. Fluxless solder reflowing is also an environmentally benign technology. The most commonly used Pb-free solder material for fluxless bonding is an alloy of Au-20(wt.%)Sn [7].

During reflowing, the solder alloy melts and then reacts with the substrate or the metallization of the chip to form intermetallic compounds (IMCs) at the joining interface. The brittle nature of these IMCs, as well as extensive intermetallic growth, can reduce the reliability of solder joints. In other words, interfacial phenomena may be directly related to the reliability of the solder joint in electronic packages. Therefore, knowledge of the formation of IMCs during soldering in microelectronic packaging is essential. In this paper, we report the fluxless Au-Sn flip-chip bumping technique using co-electroplated Au-Sn alloy. In addition, we examined the interfacial reactions between the Au-Sn solder and the Ni UBM for various reflow conditions.

## 2. EXPERIMENTAL PROCEDURES

Silicon wafers with a diameter of 4 inches were metallized with 0.2 $\mu$m-thick Ti and 0.8 $\mu$m-thick Cu. The Ti and Cu layers are used as an adhesion layer and interconnection layer, respectively. The metallized wafers had area array pad arrangements at a pitch of 300 $\mu$m with a rectangular pad opening having a width of 50 $\mu$m. The bumping for the flip chip devices was performed using electrolytic Ni with a thickness of 10 $\mu$m. Figure 1 shows the resulting Ni UBM (Under Bump Metallization) arrays employed in this study. The Ni UBM serves as both an adhesion layer and a diffusion barrier layer between the Cu and solder. The lateral overlap of the Ni UBM on the chip passivation layer was approximately 10 $\mu$m. Co-electroplating of Au-Sn alloy was performed on the Ni UBM. A commercially available single plating solution was used for the deposition of the Au-Sn alloy. The cathode was an Si wafer electroplated with Ni, and Pt was used as the anode. The electroplating cell was set up with the cathode facing the anode and the two spaced 90 mm apart. A mechanical stirrer with a controllable speed was used to supply the agitation in the electroplating solution. The temperature of the solution was controlled by means of a heater situated underneath the electroplating tank, while a thermometer placed inside the electroplating solution continuously read the temperature. A 50 $\mu$m layer of Au-Sn was electroplated in a bath at a current density of 0.5A/dm$^2$ for 120 minutes. The plating bath temperature and pH value were 35℃ and 4.3, respectively. After electroplating, the microstructural features of all of the electroplated samples were examined using a scanning electron microscopy (SEM, Philips XL 40 FEG and/or HITACHI S-3000H), equipped with an energy dispersive x-ray (EDX) spectroscope.

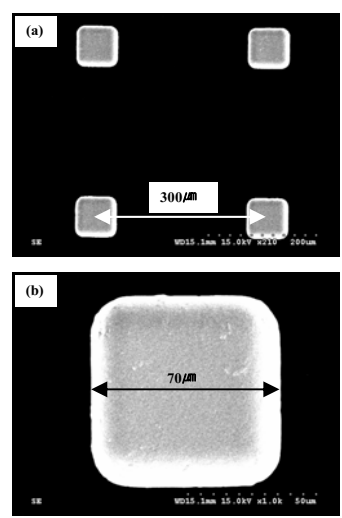

Fig.1 SEM images of Ni UBMs on a wafer.

In addition, X-ray diffraction (XRD) phase analysis was carried out using a Rigaku (Japan) diffractometer. A Cu coupon was electroplated under the same conditions for the XRD analysis. The filament voltage and current were set to 30kV and 100mA, respectively. The sample was scanned between 10° and 90° at a rate of 5°/min. Then, the electroplated Au-Sn samples were reflowed in a reflow machine (RF-430-N2, Japan Pulse Laboratory Co. Ltd., Japan) with a maximum temperature of 300℃ for 60 sec. Reflows were conducted consecutively for between one and five times. In addition, we performed reflowing to isothermally observe the morphological changes in the joint interface at 400℃ for periods ranging from 2min to 20min. Subsequently, the samples were mounted in epoxy and metallographically polished for microstructural characterization. Their microstructures and chemical compositions were observed with a SEM equipped with an EDX system. Also, the compositions of the phases formed at the interface were determined using a JEOL JXA-8900R (Tokyo, Japan) electron probe micro analyzer (EPMA) equipped with a wavelength-dispersive X-ray (WDX) analyzer. For each compositional analysis, at least five measurements were performed and the average value was reported. The total area of the interfacial IMC layer was measured using image analysis software. The layer areas were divided by the interface length shown in the cross-section to yield the average layer thickness.

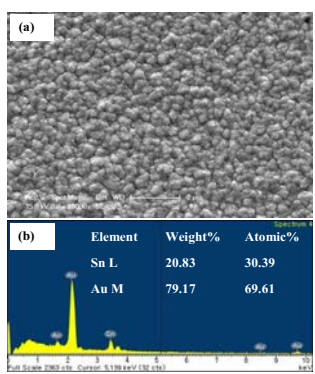

Fig.2 (a) SEM plan view image of electroplated Au-Sn alloy and (b) corresponding EDX analysis result.

## 3. RESULTS AND DISCUSSION

Au-Sn alloy electroplating was directly performed on the Ni UBM of the Si wafer. Figure 2 shows the SEM plan view image of the electroplated Au-Sn alloy and the corresponding EDX analysis result. The surface of the plating layer had a gray and rough grain structure. According to the EDX analysis result, the composition of the electroplated Au-Sn alloy was approximately 70at.% Au and 30at.% Sn. Unlike in conventional alternate electroplating, the method of co-electroplating of Au-Sn solder used in this study allowed for the direct deposition of the desired Au-Sn alloy.

Figure 3 shows the diffraction pattern obtained from the Au-Sn alloy electroplated on the Cu coupon. As expected, the alloy is a mixture of the two phases, AuSn and $Au_5Sn$. The Au-Sn eutectic alloy consists of the $Au_5Sn$ (ζ-phase) and AuSn (δ-phase) phases, as shown in Fig. 4. Using a co-electroplating process, it was possible to plate the Au-Sn solder alloy directly onto the wafer at or near the eutectic composition from a single solution. From Figs. 2 and 3, it was verified that the appropriate combination of these two phases ($Au_5Sn$ and AuSn) resulted in the eutectic composition of the Au-Sn alloy.

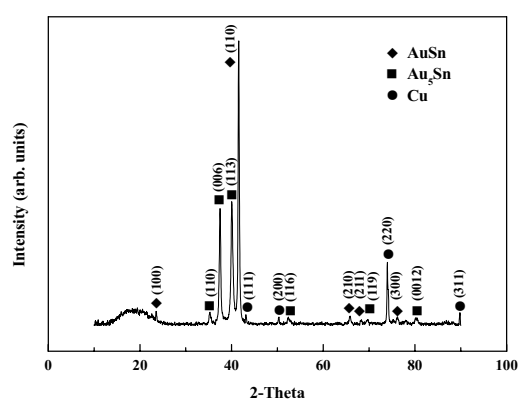

Fig.3 XRD spectra from electroplated Au-Sn sample.

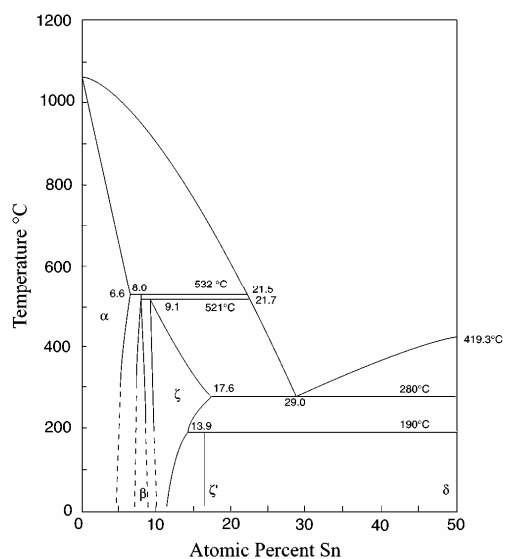

Fig.4 Au-rich portion of the Au-Sn phase diagram.

According to the Au-Sn phase diagram (Fig. 4), the eutectic temperature of the composition is 280℃. In practice, however, heating above 300℃ is necessary for the complete melting of the alloy, because the gradient of the solid-liquid line at the eutectic composition of Au-20Sn (in wt.%) is very steep.

The Au-Sn flip-chip joints are formed at 300℃ without using any flux. Figure 5 shows the SEM images of the Au-Sn solder bumps reflowed at 300℃ for 60sec. After reflowing, the average diameter of the solder bump was approximately 150 μm. The resulting Au-Sn solder bumps were smooth and exhibited a metallic silver color.

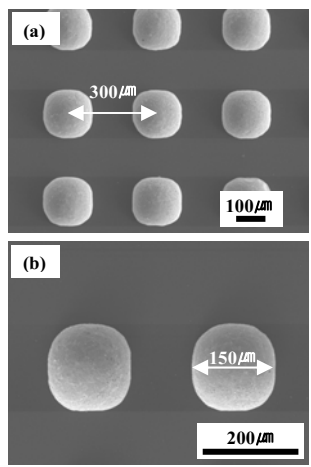

Fig.5 SEM images of the Au-Sn solder bumps reflowed on a wafer.

The typical cross-sectional microstructure of the Au-Sn solder is shown in Fig. 6. The microstructure of the solder matrix was composed of AuSn (δ-phase) and $Au_5Sn$ (ζ-phase). The bright constituent in the eutectic microstructure is the ζ-phase (nominally $Au_5Sn$) while the darker constituent is the δ-phase (AuSn).

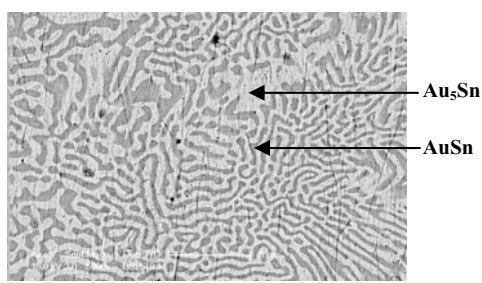

Fig.6 Cross-sectional SEM image of the Au-Sn solder reflowed at 300℃ for 60sec. The lighter phase is $Au_5Sn$ and the darker phase is AuSn.

Figure 7 shows the cross-sectional SEM images of the interfaces between the Au-Sn solder and Ni UBM after multiple reflows. The high magnification BSE (Back Scattered Electron) image mode of SEM was used to clarify the morphology of the phase formed at the interface. Only one irregular-shaped reaction product was found between the Au-Sn solder and the Ni UBM after one reflow (Fig. 7(a)). The composition of the reaction product was 25.7at.%Au-33.5at.%Ni-40.8at.%Sn. The ratio of the atomic percentage of (Au + Ni) to that of Sn was (25.7 + 33.5):(40.8), which is close to 3:2. Therefore, it is suggested that the reaction product was the $(Au,Ni)_3Sn_2$ IMC. It is known that some binary phases such as AuSn, $Ni_3Sn_4$ and $Ni_3Sn_2$ in the Au-Ni-Sn system have a very high solubility of the third element, due to the similarity in the chemical and physical properties of Au and Ni [2]. It seems that Au enters into the $Ni_3Sn_2$ lattice and substitutes for the Ni atoms. A ternary IMC often has a lower Gibbs free energy than a binary compound of the same structure from the entropy argument [2]. Therefore, $Ni_3Sn_2$ has a natural tendency to absorb Au to reach its saturated composition. A similar Ni-seeking mechanism has been proposed and widely accepted in the literature for the formation of the $(Au,Ni)Sn$ phase and/or the resettlement of the $(Au,Ni)Sn_4$ phase [2,8]. The average thickness of the $(Au,Ni)_3Sn_2$ IMC layer formed at the interface was approximately 0.6 μm. Besides the thin $(Au,Ni)_3Sn_2$ layer, rod-shape $(Au,Ni)_3Sn_2$ phases were found as well. The rod-shape $(Au,Ni)_3Sn_2$ phase had a higher Au content, as compared to the interfacial thin $(Au,Ni)_3Sn_2$ layer (see Fig. 7(a)). This phase was analyzed and found to be composed of 32.4at.% Au, 26.8at.% Ni and 40.8at.% Sn. Since Ni originates from the UBM (or substrate) and Au from the solder, it is reasonable for the Ni content of the interfacial $(Au,Ni)_3Sn_2$ layer to be higher, since it is closer to the Ni UBM.

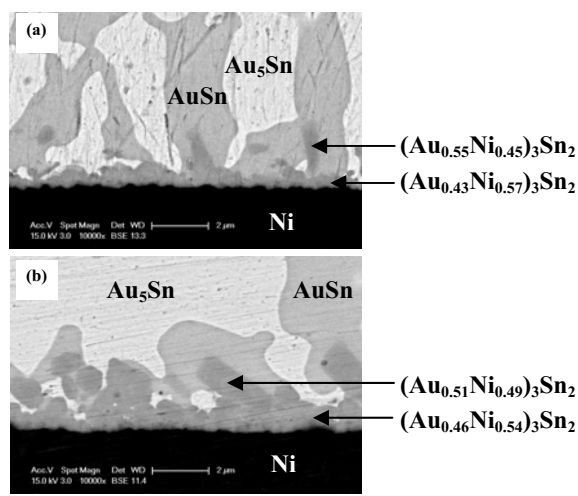

Fig.7 Cross-sectional SEM images of the Au-Sn solder/Ni UBM interfaces reflowed at 300℃ for 60sec; (a) 1 reflow and (b) 5 reflows.

Figure 7(b) shows the cross-sectional SEM image of the interface between the Au-Sn solder and Ni UBM after 5 reflows. Although the thickness of the interfacial $(Au,Ni)_3Sn_2$ IMC layer increased with increasing number of reflows, the interfacial structure was very similar to that of the single-reflow sample. In addition, many IMC particles were found at the interface. However, these IMC particles were actually cross-sections of the rod-shape IMCs grown randomly from different locations on the interface.

Figure 8 shows the cross-sectional SEM images of the Au-Sn solder/Ni UBM interfaces reflowed at 400℃ for different reaction times. After reflowing at 400℃ for 2 min, two IMC layers, $(Au,Ni)_3Sn_2$ and $(Au,Ni)_3Sn$, were formed at the interface, as shown in Fig. 8(b). The $(Au,Ni)_3Sn_2$ phase is initially formed when the liquid Au-Sn solder reacts with the Ni UBM, and then the $(Au,Ni)_3Sn$ phase is formed when the $(Au,Ni)_3Sn_2$ phase reacts with the Ni UBM. The thickness of the $(Au,Ni)_3Sn$ layer formed on the Ni UBM was very thin. In addition, $(Au,Ni)_3Sn_2$ IMC particles were widely dispersed in the matrix of the solder alloy (see Fig. 8(a)). As a whole, the thickness of the $(Au,Ni)_3Sn_2$ and $(Au,Ni)_3Sn$ IMC layers increased with increasing reaction time, as shown in Fig. 8. After reflowing at 400℃ for 20 min, the thickness of the $(Au,Ni)_3Sn_2$ and $(Au,Ni)_3Sn$ IMC layers were about 2.1 and 0.4 $\mu m$, respectively. The measured average compositions of the $(Au,Ni)_3Sn_2$ and $(Au,Ni)_3Sn$ layers in Fig. 8(f) are shown in Table 1. The composition varied across the upper $(Au,Ni)_3Sn_2$ layer with the Au content being slightly higher on the solder side and the Ni content increasing toward the UBM.

In addition, the microstructure inside the solder was lamellar and was composed of AuSn (δ-phase) and $Au_5Sn$ (ζ-phase) (see Figs. 8(a) and (c)). As the reflow time increased, the eutectic lamellae in the bulk solder coarsened as shown in Fig. 8 (e).

Table 1 Chemical compositions (in at.%) of the IMC phases formed at the interface of Fig. 8.(f)

| Phases | Au | Ni | Sn |
|---|---|---|---|
| $(Au,Ni)_3Sn_2$ | 15.9 | 43.8 | 40.3 |
| $(Au,Ni)_3Sn$ | 4.9 | 72.5 | 22.6 |

A peculiar phenomenon was observed in the reflowed solder matrix. Figure 9 shows the cross-sectional SEM images of the solder matrix in the Au-Sn solder/Ni UBM joint reflowed at 400℃ for 2 min. As shown in Fig. 9, the spalled $(Au,Ni)_3Sn_2$ phase was embedded in the AuSn phase. The dark core region is $(Au,Ni)_3Sn_2$ phase, and the light exterior layer surrounding the $(Au,Ni)_3Sn_2$ phase is AuSn (δ-phase). The measured average compositions of the $(Au,Ni)_3Sn_2$ and AuSn phases are shown in Table 2. Similar phenomena are also observed in the interfacial SEM images of the samples reflowed at 300 and 400℃ (see Figs. 7 and 8). As the temperature and time of the reflow increased, the δ(AuSn)-phase adjacent to the interfacial layer was gradually replaced by the ζ($Au_5Sn$)-phase, as shown in Figs. 7(a), 7(b) and 8(e). Eventually, the ζ-phase covered mainly the interfacial layer, as shown in Fig. 8(e). In other words, the phase distribution at the interface changed by the interfacial reaction proceeded, due to reflowing.

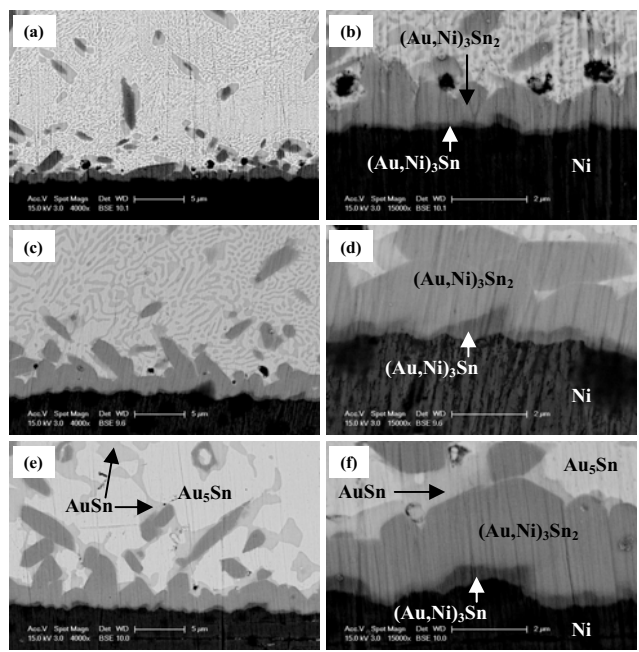

Fig.8 Cross-sectional SEM images of the Au-Sn solder/Ni UBM interfaces reflowed at 400℃ for (a and b) 2 min, (c and d) 10 min and (e and f) 20 min.

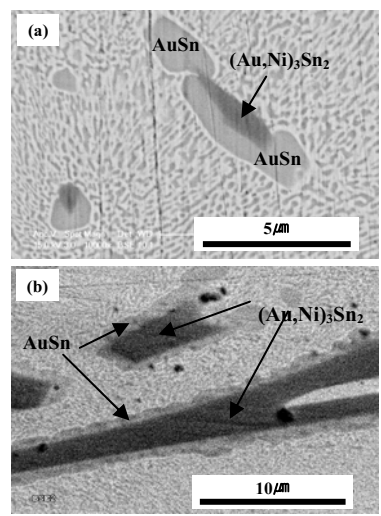

Fig.9 SEM images of the solder matrix in the Au-Sn solder/Ni UBM joint reflowed at 400℃ for 2 min.

Table 2 Chemical compositions (in at.%) of the phases indicated in Fig. 9.

| Phases | Au | Ni | Sn |
|---|---|---|---|
| $(Au,Ni)_3Sn_2$ | 17.3 | 40.1 | 42.6 |
| AuSn | 51.2 | - | 48.8 |

## 4. CONCLUSION

In this study, we fabricated eutectic or near eutectic Au-Sn flip-chip bumps from a single plating bath using the principle of alloy co-electroplating. This process is well suited to the deposition of Au-Sn alloy with a composition of Au-30at.%Sn. Fluxless soldering can be performed with the plated Au-Sn solder bump. After reflowing, the average diameter of the solder bump was approximately 150 $\mu$m. The microstructure of the solder matrix was composed of the eutectic phases of the Au-Sn alloy, AuSn ($\delta$-phase) and $Au_5Sn$ ($\zeta$-phase). We also studied the initial microstructure and microstructural evolution of the eutectic Au-Sn solder bumps on the Ni UBM during the reflow reaction. In the case of the samples reflowed at 300℃, only an $(Au,Ni)_3Sn_2$ IMC layer formed at the interface between the Au-Sn solder and Ni UBM. On the other hand, two IMC layers, $(Au,Ni)_3Sn_2$ and $(Au,Ni)_3Sn$, were found at the interfaces of the samples reflowed at 400℃. The $(Au,Ni)_3Sn_2$ phase is initially formed when the liquid Au-Sn solder reacts with the Ni UBM, and then the $(Au,Ni)_3Sn$ phase is formed when the $(Au,Ni)_3Sn_2$ phase reacts with the Ni substrate. As the reflow time increased, the thickness of the interfacial $(Au,Ni)_3Sn_2$ and $(Au,Ni)_3Sn$ IMC layers increased.

## 5. ACKNOWLEDGEMENT

This work was supported by grant No. RTI04-03-04 from the Regional Technology Innovation Program of the Ministry of Commerce, Industry and Energy(MOCIE).

## 6. REFERENCES

[1] J.W.R. Tew, X.Q. Shi, and S. Yuan, "Au/Sn solder for face-down bonding of AlGaAs/GaAs ridge waveguide laser diodes," *Materials Letters*, Elsevier, pp. 2695-2699, 2004.

[2] J.Y. Tsai, C.W. Chang, Y.C. Shieh, Y.C. Hu, and C.R. Kao, "Controlling the microstructures from the gold-tin reaction," *Journal of Electronic Materials*, TMS, pp. 182-187, 2005.

[3] J. Doesburg and D.G. Ivey, "Microstructure and preferred orientation of Au-Sn alloy plated deposits," *Materials Science and Engineering B*, Elsevier, pp. 44-52, 2000.

[4] B. Djurfors and D.G. Ivey, "Pulsed electrodeposition of the eutectic Au/Sn solder for optoelectronic packaging," *Journal of Electronic Materials*, TMS, pp. 1249-1254, 2001.

[5] B. Djurfors and D.G. Ivey, "Microstructural characterization of pulsed electrodeposited Au/Sn alloy thin films," *Materials Science and Engineering B*, Elsevier, pp. 309-320, 2002.

[6] D.W. Kim and C.C. Lee, "Fluxless flip-chip Sn-Au solder interconnect on thin Si wafers and Cu laminated polyimide films," *Materials Science and Engineering A*, Elsevier, pp. 74-79, 2006.

[7] G. Elger, M. Hutter, H. Oppermann, R. Aschenbrenner, H. Reichl, and E. Jager, "Development of an assembly process and reliability investigations for flip-chip LEDs using AuSn soldering," *Microsystem Technologies*, Springer, pp. 239-243, 2002.

[8] K.Y. Lee, M. Li, and K.N. Tu, "Growth and ripening of $(Au,Ni)Sn_4$ phase in Pb-free and Pb-containing solders on Ni/Au metallization," *Journal of Materials Research*, MRS, pp. 2562-2570, 2003.